\documentclass[a4paper,11pt]{article}
\usepackage{jinstpub} 
\usepackage{siunitx}

\proceeding{21$^{\text{st}}$ ``Trento'' Workshop on Advanced Silicon Radiation Detectors\\
February 2026\\
Perugia, Italy}

\title{\boldmath Novel Strip-like Readout Geometries in Resistive AC-coupled Silicon Detectors (RSD / AC-LGAD)}

\author[a,c]{L.L. Grimm,\note{Corresponding author.}}
\author[a]{L. Hahn,}\author[a,1]{B. Regnery,}\author[b,c]{L. Menzio,}\author[b,d]{R. Arcidiacono,}\author[b]{N. Cartiglia,}\author[a]{A. Dierlamm,}\author[a]{M. Klute,}\author[c]{and M. Moll}
\affiliation[a]{Institute of Experimental Particle Physics (ETP), Karlsruhe Institute of Technology (KIT), 76131 Karlsruhe, Germany}
\affiliation[b]{Sezione di Torino, Istituto Nazionale di Fisica Nucleare, Via Pietro Giuria 1, 10125 Torino, Italy}
\affiliation[c]{CERN, European Organization for Nuclear Research,
Esplanade des Particules 1, Geneva, 1211, Switzerland}
\affiliation[d]{Dipartimento di Scienze del Farmaco, Universita del Piemonte Orientale, Largo Donegani 2, 28100 Novara, Italy}

\emailAdd{brendan.regnery@kit.edu}

\abstract{
Resistive Silicon Detectors (RSD/AC-LGAD) are novel silicon detectors capable of both precise spatial and temporal resolution. Such sensors will be essential for the next generation of particle colliders (EIC, FCC-ee, CEPC, FCC-hh) and would enable the possibility of a 4D tracker. RSD sensors are typically fabricated with a pixel-like geometry that provides excellent spatial resolution in the x and y directions. However, in regions further from the interaction point, high spatial resolution in one direction (strip-like geometry) is often preferred to reduce the number of readout channels. For example, strip AC-LGADs are now the default option for the US electron ion collider (EIC). The second production of RSD sensors by Fondazione Bruno Kessler includes sensors with unconventional readout pad shapes that act as a mixture between strip-like and pixel-like readout. This work presents the first characterization of these new pad designs using the Transient Current Technique (TCT). The measurements demonstrate exceptional one-dimensional spatial resolution, confirming the potential of novel strip-like RSDs for future tracking systems.}


\arxivnumber{} 

\begin{document}
\maketitle
\flushbottom

\section{Introduction}
\label{sec:intro}
The field of high-energy physics is rapidly advancing towards a next generation of particle detectors that will navigate high pile-up (e.g. HL-LHC \cite{HL-LHC}, FCC-hh \cite{FCChh}) and precision particle identification (e.g. FCC-ee \cite{FCCee}, CEPC \cite{CEPC}). In all cases, precise timing detectors will be vital for any future collider. Recent developments have resulted in sensors that utilize internal gain to provide precise temporal resolution, Low Gain Avalanche Diodes (LGADs) \cite{UFSDpaper, NicoloBook}.

The HL-LHC upgrades of the CMS and ATLAS experiments will use LGADs with a pitch of roughly $1.3\times1.3$~\si{\milli\meter\squared} and a temporal resolution $<$ \SI{50}{\pico\second} \cite{ATLAS_HGTD, CMS_MTD}. These sensors will be exploited in timing layers to provide tags for individual tracks and ultimately improve pile-up mitigation in high rate environments. In recent years, the LGAD technology has been continuing to develop into sensors that provide both precise spatial and temporal resolution. In a future lepton collider, such sensors could be used in a precision time-of-flight detector that would become an integral part of any particle identification system. In a future hadron collider, sensors with hits in space and time could be used to create a `4D’ tracking system that would greatly reduce the track finding combinatorics.

\begin{figure}[htbp]
    \centering
    \includegraphics[width=.8\textwidth]{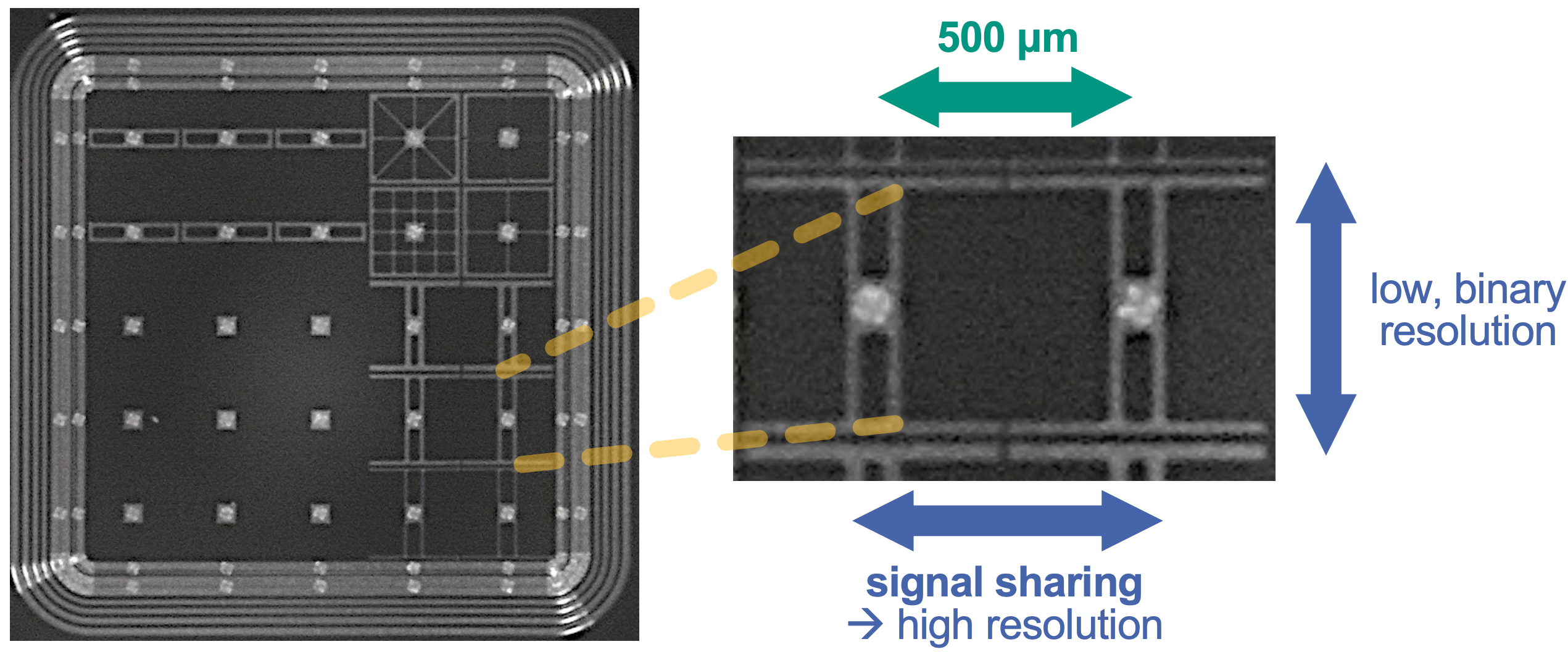}
    \caption{An RSD sensor from the second FBK production containing various possible electrode shapes. The zoomed in region highlights the H-shaped, hybrid pixel-strip electrode shapes used in this study. \label{fig:sensor}}
\end{figure}

One such novel sensor is the Resistive Silicon Detector (RSD/AC-LGAD) which, like LGADs, uses a gain implant for time resolution but differs by including a continuous resistive n+ layer. This results in an AC-coupled, n-in-p sensor with electrodes on an oxide layer placed above the n$^{+}$ and p$^{+}$ layers. The sensor is biased from the backside with a path to ground provided via a bias ring contacting the n+ layer. Passing charged particles induce a signal which is shared between multiple electrodes \cite{RSD1MarcoM}. An example of such readout electrodes is shown in figure~\ref{fig:sensor} and the sensor structure can be seen in the next section in figure~\ref{fig:TCT}. This approach allows for an innate 100\% fill factor and precise spatial reconstruction with large electrode pitches. Ultimately, this results in fewer readout channels, requiring readout chips with lower power consumption and, thus, less cooling \cite{NicoloBook, RSD1Marta}.

RSD sensors are typically fabricated with pixel-like geometry for excellent spatial resolution in both $x$- and $y$-direction \cite{rsdPosition, SivieroPosition}. However, in regions further from the interaction point, one dimensional spatial resolution is often enough so that sensor cost, readout electronics, and cooling can be even further reduced. The second production of RSDs from Fondazione Bruno Kessler (FBK) included sensors with various electrode designs, including several in the shape of an H \cite{RSDsecondProd}. This geometry was designed to provide charge sharing in one direction, creating a hybrid strip-pixel geometry. The studies reported in this article aim to measure the one dimensional spatial resolution with the Transient Current Technique for these novel RSD designs.

The hybrid pixel-strip RSD sensor consists of an H-shaped electrode with a pitch of \SI{500}{\micro\meter}. The resulting geometry is shown in figure~\ref{fig:sensor} along with other potential electrode shapes. The H-shapes were designed with the intention of limiting the charge spread to only two electrodes. Thus, this hybrid design should provide precise one dimensional spatial resolution between the cross-bars of two H-shapes (i.e. the zoomed-in region in figure~\ref{fig:sensor}), while neighboring H-shapes will have limited improvement from charge sharing, resulting in a more traditional strip/pixel resolution. 

\section{Measurement with the Transient Current Technique (TCT)}

\begin{figure}[htbp]
\centering
\includegraphics[width=.7\textwidth]{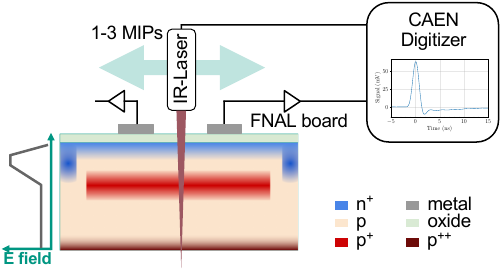}
\caption{Diagram of the Top-TCT system at KIT. The RSD sensor was read out with a 16-channel FNAL LGAD amplification board connected to a CAEN digitizer with 5~GS/s sampling rate. \label{fig:TCT}}
\end{figure}

The hybrid pixel-strip RSD sensor was studied using the Transient Current Technique (TCT) system at KIT. The setup utilized a large-area Top-TCT system consisting of a \SI{1055}{\nano\meter} pulsed laser with a spot size of \SI{15}{\micro\meter} FWHM guided by a 3D translation stage. For this study, the signal was read out by a 16-channel Fermilab (FNAL) LGAD amplification board connected to a CAEN digitizer with a 5~GS/s sampling rate. In order to study particle-like response, the laser intensity was adjusted to create signals comparable to 1-3 MIPs. The MIP signal strength was quantified using signals created by electrons emitted by a Sr-90 source. A diagram of the TCT setup is shown in figure~\ref{fig:TCT}. 

During measurements, the laser spot was scanned over the sensor area in a grid with \SI{25}{\micro\meter} step size and the signal was recorded for 100 events at each position. The scan was repeated for bias voltages over a range of \SI{30}{\volt} to \SI{230}{\volt} in \SI{10}{\volt} increments.

The signal was corrected by the baseline noise and the waveforms were averaged at each position. From this, the signal amplitude was then used for further analysis. Points where the signal is reduced due to wire bonds or metallization of the sensor were excluded from the analysis. Due to differences in channel amplification on the FNAL readout board, the signal is normalized. This improves comparability between different pairs of electrodes, which is vital for hit reconstruction.

\begin{figure}[htbp]
\centering
\includegraphics[width=.8\textwidth]{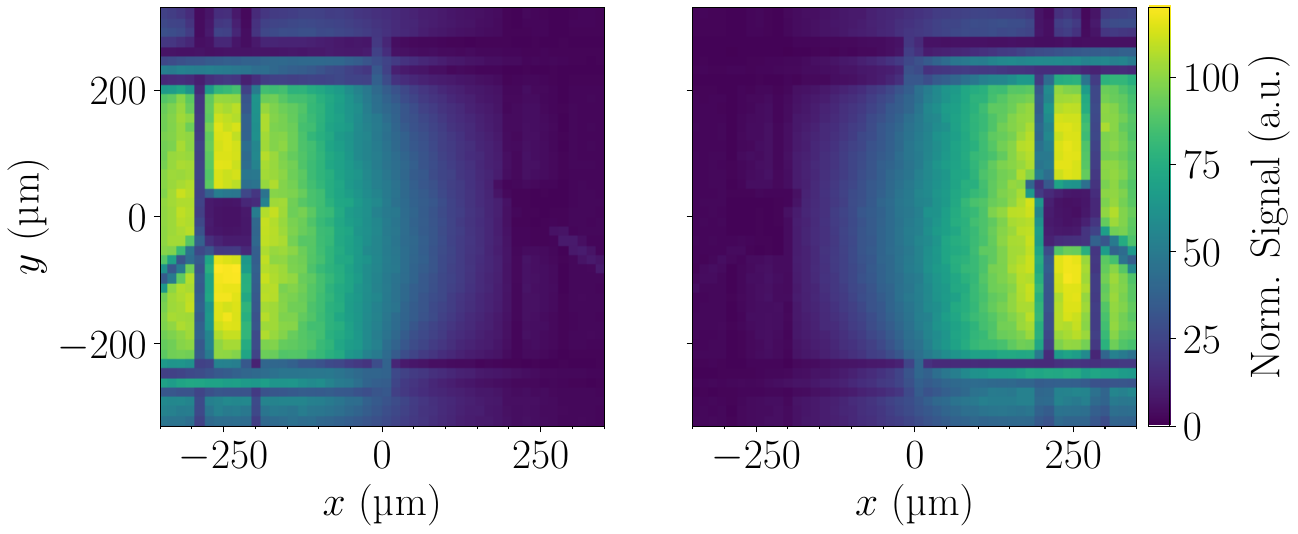}
\caption{2D map of the signal amplitude on the left and right electrode obtained from scanning the laser over the sensor at $V_{\rm{bias}}=210$~\si{\volt}. \label{fig:2dmap}}
\end{figure}

Figure~\ref{fig:2dmap} shows a 2D map of the scan for the left electrode at $V_{\rm{bias}}=210~$\si{\volt}. The bluish shadows indicate regions where the wire-bonds or metalized electrodes block the laser. The following discussion focuses on the region between the electrodes. A consistent signal decrease along the $x$-axis is observed across different $y$-positions, as evidenced by the projections in Figure~\ref{fig:projection}. The spread between these curves is small compared to the signal, suggesting a high degree of uniformity. Furthermore, the signal profile appears linear, motivating the following reconstruction method.

\section{Reconstruction Method}

As electron-hole pairs are created in the RSD sensor, the electrons will follow the path of least resistance and induce a signal on the electrodes. Thus, the simplest reconstruction technique used in RSD sensors is to treat the sensor as a network of resistors and perform a weighted fit between the electrodes.

\begin{figure}[htbp]
\centering
\includegraphics[width=.8\textwidth]{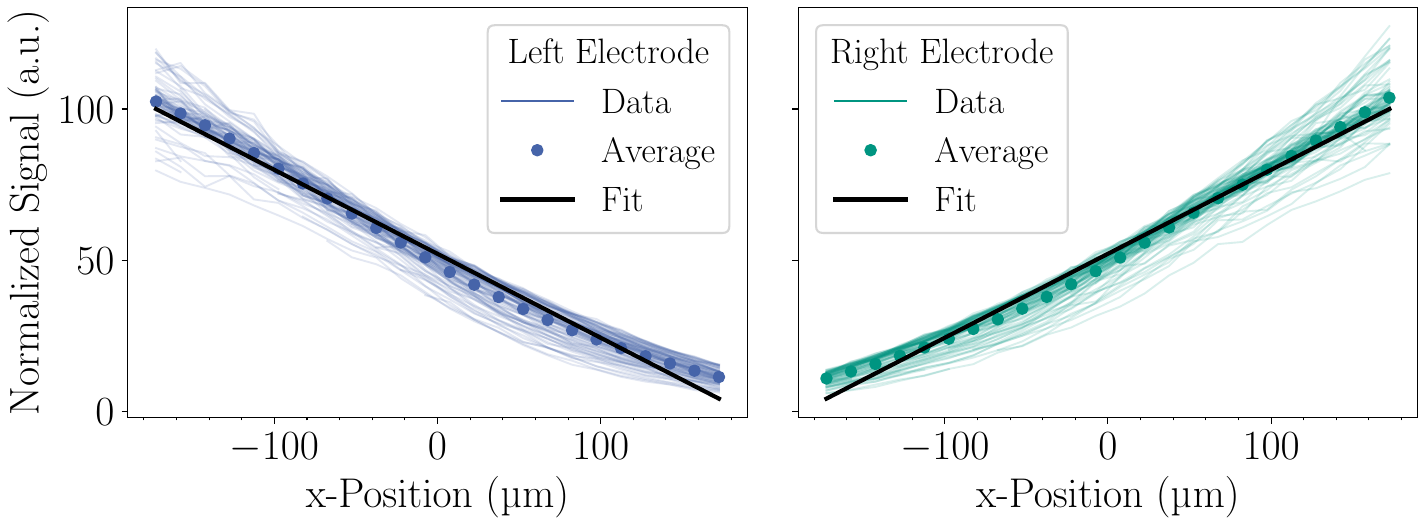}
\caption{Projections of the signal amplitude along $x$- for various $y$-positions for the left and right electrode. The lightly colored lines are projections for a given $y$-position, the points denote the average for all $y$-positions, and the line is the resulting linear fit. \label{fig:projection}}
\end{figure}

Figure~\ref{fig:projection} shows one dimensional projections of the signal amplitude along $x$ for various $y$-positions for two electrodes (i.e. one dimensional projections of the plots in figure~\ref{fig:2dmap}). The projections follow a linear decrease as a function of distance from the electrode. This decrease is quantified by averaging the projections. Then, the signal amplitude versus $x$ is modeled by performing a linear fit to the average projection
\begin{equation}
    S_{\rm{fit}}(x) = a x + b \,.
\end{equation} 
The results of the projections, average, and linear fit for the left and right electrode are shown in figure~\ref{fig:projection}. 

Finally, the reconstructed hit position is determined using the results of the linear fits to both electrodes. This is done by using the recorded signal amplitude $S_{\rm{meas}}$ and fit results $S_{\rm{fit}}(x)$ to construct a $\chi^{2}$ function with 
\begin{equation}
    \chi^2(x) = (S_{\rm{meas, left}} - S_{\rm{fit, left}}(x))^2 + (S_{\rm{meas, right}} - S_{\rm{fit, right}}(x))^2 \, .
\end{equation}
The hit position is then the $x$-value which minimizes this $\chi^{2}$.

\section{Results}

\begin{figure}[htbp]
\centering
\includegraphics[width=.4\textwidth]{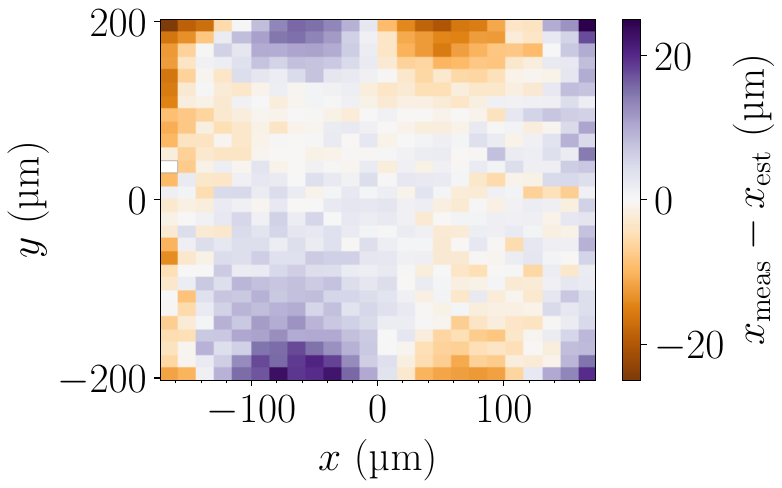}
\qquad
\includegraphics[width=.4\textwidth]{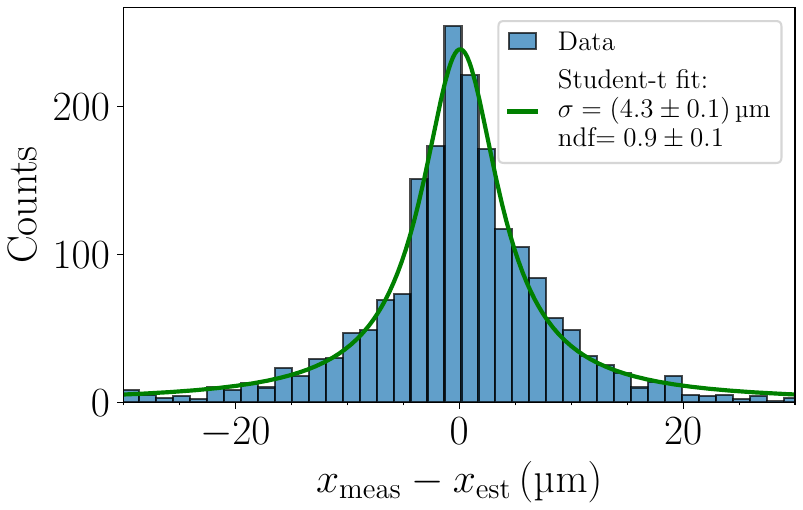}
\caption{(left) The difference between the position of the laser ($x_{\rm{meas}}$) and the reconstructed hit position ($x_{\rm{est}}$) for each $(x,y)$ point between the electrodes. Blue indicates regions where the hit was reconstructed left of the laser while orange regions indicate a hit reconstructed to the right of the laser. The white dot corresponds a region excluded from the analysis where the laser was obstructed due to a wire bond. (right) A histogram of the differences fit with a Student-t distribution. Both plots are for measurements at $V_{\rm{bias}}=210$~\si{\volt}. \label{fig:residuals}}
\end{figure}

First, the measured position of the laser ($x_{\rm{meas}}$) was compared to the reconstructed hit position ($x_{\rm{est}}$) and then the difference was plotted for each point along the sensor. This relative offset is shown in figure~\ref{fig:residuals} (left). Blue regions indicate where the estimated position is shifted left of the laser, while orange regions indicate a shift to the right. The center region between the electrodes has the most accurate reconstruction with a degradation near the electrode edges.

Then, the spatial resolution is extracted by plotting the differences in a histogram and fitting with a Student-t distribution. An example of the fit is shown in figure~\ref{fig:residuals} (right); the width ($\sigma$) of the Student-t corresponds to the spatial resolution.

\begin{figure}[htbp]
\centering
\includegraphics[width=.6\textwidth]{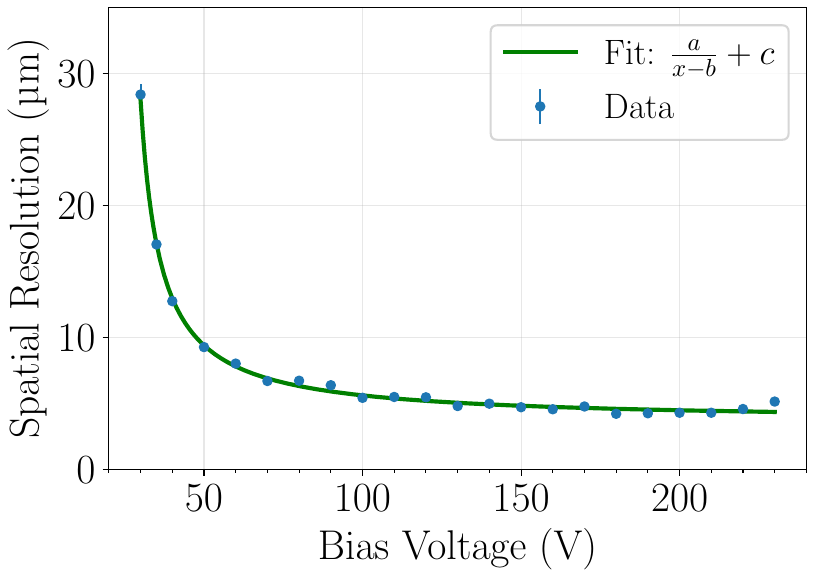}
\caption{Spatial resolution as a function of bias voltage. The plateau at \SI{5}{\micro\meter} corresponds to $\sim 1$\% of the sensor pitch. \label{fig:bias_scan}}
\end{figure}

The resulting spatial resolution as a function of bias voltage is shown in figure~\ref{fig:bias_scan}. The resolution improves with increasing bias voltage due to the higher signal-to-noise ratio (increasing exponential from 1 to 60), approaching a plateau of approximately \SI{5}{\micro\meter}. The achieved spatial resolution corresponds to $\sim1$\% of the \SI{500}{\micro\meter} electrode pitch, demonstrating the effectiveness of resistive charge sharing.

\section{Conclusions}

RSDs are promising silicon sensors which provide precise spatial and temporal resolution. A new hybrid pixel-strip geometry was studied using Top-TCT to obtain the one dimensional spatial resolution. These sensors achieved a spatial resolution of $\sim1$\% of the sensor pitch using a simple, weighted reconstruction. Obtaining such results with 2D RSD sensors requires constructing and using detailed template lookup tables or utilizing complex machine learning \cite{rsdPosition, SivieroPosition}. The precise spatial resolution began to plateau at relatively low bias voltages, a result previously unobserved. In the future, the temporal resolution of these sensors will be verified. 

The significance of this result lies in the simplicity. The simple reconstruction technique could be used by circuit designers to reduce the complexity needed in the readout circuitry. Such a reduction leads to less power consumption and less cooling demands. The electrode shapes could be used for hybrid sensors aiming to optimize readout in one-dimension and potentially be expanded to create a longer chain of `strip-like’ electrodes that could be used to further improve the one dimensional spatial resolution of current AC-LGAD strip sensors \cite{ACLGADstrip}. A precise, strip-like RSD could be used to create modules where readout circuitry is placed on the sensor periphery and connected via wire-bonds, further reducing the material budget. However, further studies are needed into time-resolution after signal propagation along such strips. Such design choices will be critical during the creation of a precision lepton collider and make strip-like RSD geometries a crucial component of future sensor development.


\acknowledgments

The authors would like to acknowledge the support of the following funding agencies and collaborations: Alexander von Humboldt Stiftung; KIT KCETA Ausschreibung Sachmittel; INFN - CSN5 RSD project; Dipartimenti di Eccellenza, Univ. of Torino (ex L. 232/2016, art. 1, cc. 314, 337); European Union - Next Generation EU, Mission 4 component 2, CUP C53D23001510006. The authors would like to thank Hans Jürgen Simonis and Tobias Barvich for technical support and Lea Stockmeier for reviewing the manuscript. 





\bibliographystyle{JHEP}
\bibliography{biblio.bib}
\end{document}